\documentclass[aps,prb,twocolumn,superscriptaddress,longbibliography]{revtex4-1}
\usepackage{amsmath, amssymb}
\usepackage{graphicx}
\usepackage{hyperref}
\usepackage{color}
\usepackage{dcolumn} 

\usepackage[english]{babel}
\usepackage{letltxmacro}

\LetLtxMacro{\ORIGselectlanguage}{\selectlanguage}
\makeatletter
\DeclareRobustCommand{\selectlanguage}[1]{%
  \@ifundefined{alias@\string#1}
    {\ORIGselectlanguage{#1}}
    {\begingroup\edef\x{\endgroup
       \noexpand\ORIGselectlanguage{\@nameuse{alias@#1}}}\x}%
}
\newcommand{\definelanguagealias}[2]{%
  \@namedef{alias@#1}{#2}%
}
\makeatother

\definelanguagealias{en}{english}

\renewcommand{\vec}[1]{{\bf #1}}

\begin{document}
\title{THz conductivity of graphene on boron nitride}
\author{Ashley M. DaSilva}
\affiliation{Department of Physics, The University of Texas at Austin, Austin, Texas 78712-1192, USA}
\author{Jeil Jung}
\affiliation{Department of Physics, University of Seoul, Seoul 130-743, Korea}
\author{Shaffique Adam}
\affiliation{Centre for Advanced 2D Materials and Department of Physics, National University of Singapore, 2 Science Drive 3, Singapore 117542}
\affiliation{Yale-NUS College, 16 College Ave West, Singapore 138527}
\author{Allan H. MacDonald}
\affiliation{Department of Physics, The University of Texas at Austin, Austin, Texas 78712-1192, USA}

\begin{abstract}
The conductivity of graphene on a boron nitride substrate exhibits features in the terahertz (THz) and infrared (IR) frequency regimes that are 
associated with the periodic moir{\' e} pattern formed by the weakly coupled two-dimensional materials.  
The THz and IR features are strongest when the two honeycomb lattices are orientationally aligned, and 
in this case are Pauli blocked unless the Fermi level is close to $\pm 150$~meV relative to the
graphene sheet Dirac point.  
Because the transition energies between moir{\' e} bands formed above 
the Dirac point  are small, {\em ac} conductivity features in n-doped graphene tend to 
be overwhelmed by the Drude peak.  The substrate-induced band splitting is larger 
at energies below the Dirac point, however, and can however lead to sharp features at THz and IR 
frequencies in p-doped graphene.
In this Letter we focus on the strongest few THz and IR features, explaining how they arise 
from critical points in the moir{\' e}-band joint density-of-states, and 
commenting on the interval of Fermi energy over which they are active.  
\end{abstract}

\maketitle

\noindent
{\em Introduction:}---
Strong interactions between graphene sheets and light have motivated research aimed at potential graphene-based optoelectronic 
devices\cite{rana_graphene_2008,bonaccorso_graphene_2010,vakil_transformation_2011,sensale-rodriguez_broadband_2012,liu_gate-tunable_2014} 
operating in the terahertz (THz) and the infrared (IR) frequency regimes, a range of the electromagnetic spectrum 
that is important for imaging, sensing, and communications technologies.
\cite{hu_imaging_1995,kemp_security_2003,kawase_terahertz_2004,koch_terahertz_2007,tang_cmos_2014} 
The ability to tune carrier density in graphene using gate voltages\cite{novoselov_electric_2004} is a 
great advantage in most application possibilities. 

\begin{figure}
\includegraphics[width=1\columnwidth]{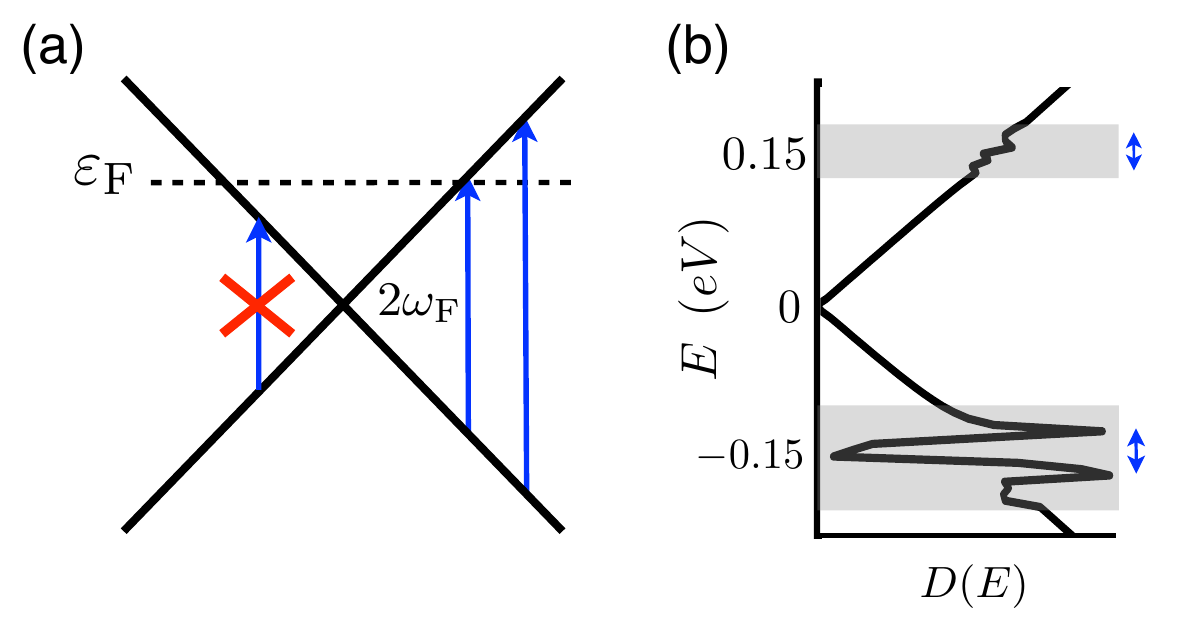}
\caption{{\bf Pauli blocking in bare graphene.} 
(a) This cartoon shows the Dirac cone of doped isolated graphene with a Fermi level $\varepsilon_{F}$ given by the dashed line. Blue arrows represent interband transitions from the valence band to the conduction band. The middle arrow, with frequency $2\omega_{F}=2\varepsilon_{F}/\hbar$, is at the minimum frequency for interband transitions. Below $2\omega_{F}$, transitions are not allowed due to the Pauli exclusion principle.
(b) Density of states of aligned graphene on hBN. 
The shaded regions represent the locations of the substrate induced gapped 
Dirac cone replicas.  The conductivity is strongly Fermi level dependent near these 
energies because of Pauli blocking.}
\label{fig:pauliBlocking}
\end{figure}

The {\em ac} conductivity of a neutral graphene sheet\cite{wang_gate-variable_2008,li_dirac_2008,ando_dynamical_2002} 
is nearly frequency-independent $\sigma_{0}=\pi e^{2}/2h$; adding n or p type carriers transfers 
oscillator strength over the frequency interval $(0,2 \omega_F)$ which is Pauli blocked for 
interband transitions to a Drude peak of equal weight.  
When a graphene sheet is aligned with a hexagonal boron nitride substrate secondary Dirac points are 
induced by the substrate and, because they are gapped, lead to a large joint density-of-states 
for low-frequency optically active transitions.  Like the interband conductivity of an isolated 
graphene sheet, the sharp conductivity features associated with these transitions can be 
turned off and on by adjusting the position of the Fermi level.    
(See Fig.~\ref{fig:pauliBlocking}).  In this Letter we present a theory of the conductivity at 
Fermi level values close to secondary Dirac points.   

\begin{figure*}
\includegraphics[width=5in]{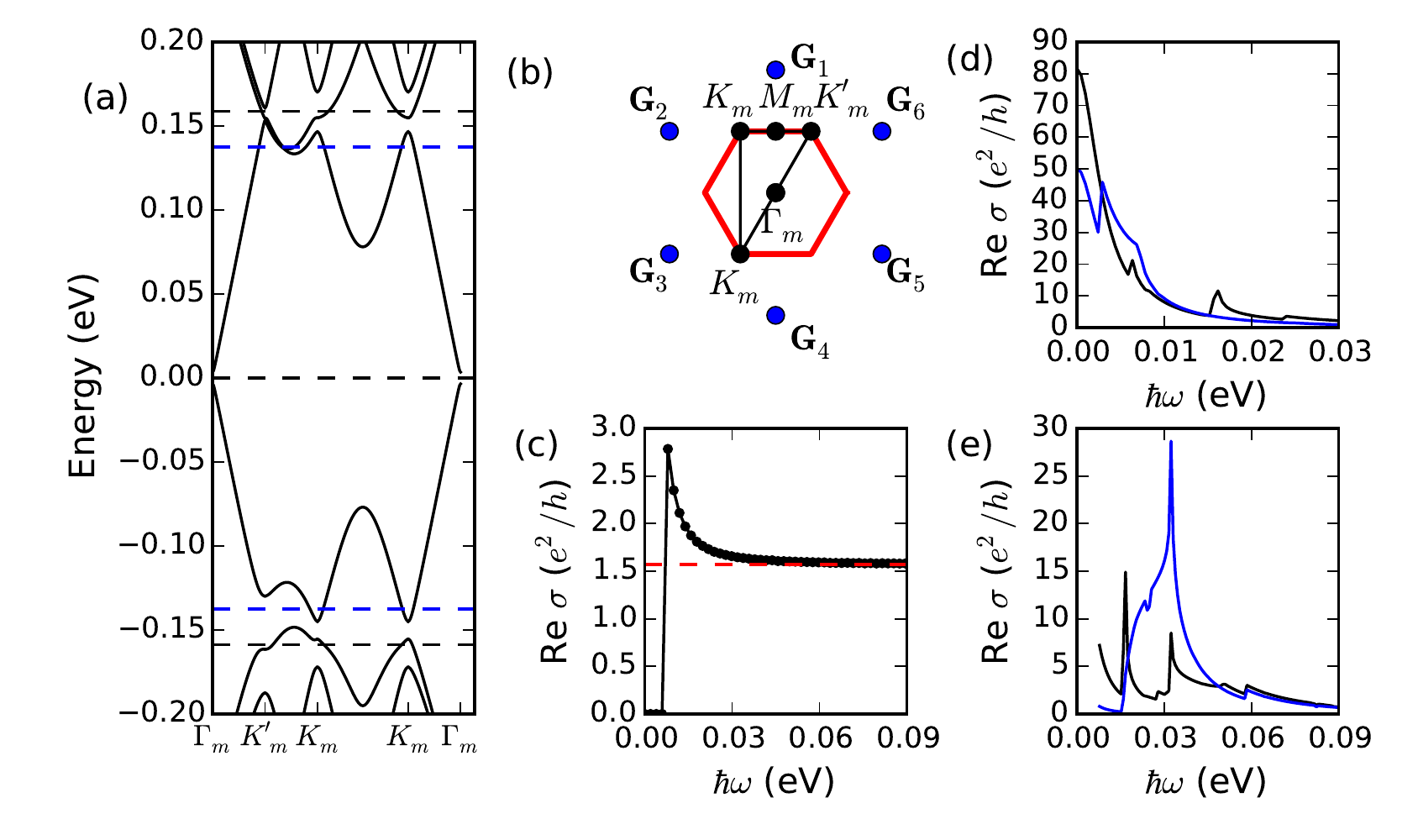}
\caption{{\bf Band structure and {\em ac} conductivity of graphene on boron nitride.} 
(a) Band structure of orientationally aligned graphene on boron nitride. 
The dashed lines show the Fermi levels for which the {\em ac} conductivity has been calculated. 
(b) Schematic picture of the moir{\'e} Brillouin zone, outlined in red. 
The bands are plotted along the black lines in momentum space. Blue dots represent the first shell of moire{\'e} reciprocal lattice vectors. 
(c) The {\em ac} conductivity when the Fermi level is at the Dirac point $\varepsilon_{F}=0$. 
The red dashed line shows that $\sigma(\omega)$ approaches the bare graphene value $\sigma_{0}=\pi e^{2}/2h$ at high 
frequencies. (d) The {\em ac} conductivity when the Fermi level is in the conduction band and near the secondary Dirac point
with $\varepsilon_{F}=\hbar v G/\sqrt{3}$ (black) and near the secondary Dirac point with $\varepsilon_{F}=\hbar v G/2$ (blue). 
$G$ is the magnitude of the moir{\'e} reciprocal lattice vectors in the first shell. 
(e) The {\em ac} conductivity when the Fermi level is in the valence band and near the secondary Dirac point
with $\varepsilon_{F}=-\hbar v G/\sqrt{3}$ (black) and $\varepsilon_{F}=-\hbar v G/2$ (blue).}
\label{fig:acconductivity}
\end{figure*}

Hexagonal boron nitride is a wide band gap semiconductor\cite{watanabe_direct-bandgap_2004} with a hexagonal lattice structure and weakly coupled layers.  
Placing graphene on a hexagonal boron nitride substrate retains the high quality of graphene\cite{dean_boron_2010,xue_scanning_2011,decker_local_2011} while modifying its band structure and 
therefore its optical response. 
The substrate's pattern of negatively charged boron atoms and positively charged nitrogen atoms on 
opposite honeycomb sublattices alters the graphene $\pi$-band Hamiltonian.
If the two honeycomb lattices had perfectly matched lattice constants and orientations, 
the primary effect of this interaction would  be to simply to open a 
gap at the Dirac point. \cite{giovannetti_substrate-induced_2007,jung_ab_2014} 
However, the boron nitride lattice constant is about two percent larger than that of graphene and differences in orientation are typical in exfoliated samples,\cite{ortix_graphene_2012,xue_scanning_2011} leading to a more complex electronic structure. 
In nearly aligned layers, long-period moir{\'e} patterns\cite{xue_scanning_2011,decker_local_2011,yankowitz_emergence_2012} 
form and influence all physical properties.   
Experiments show that a gap opens at the Fermi level of neutral graphene sheets, 
with a value that is dependent on electron-electron interactions,
and on strains induced by the lattice constant and orientation 
mismatch.\cite{hunt_massive_2013,ponomarenko_cloning_2013,woods_commensurate-incommensurate_2014}
At perfect alignment (zero relative rotation angle between the graphene and boron nitride),
the moir{\'e} wavelength is around 15 nm. 
In addition to the gap at charge neutrality, secondary gapped Dirac points 
appear\cite{park_new_2008,park_new_2008,yankowitz_emergence_2012,kindermann_zero-energy_2012,ortix_graphene_2012,wallbank_generic_2013,mucha-kruczynski_heterostructures_2013,jung_origin_2015}
at Fermi energies corresponding to $\pm 4$ electrons per moir{\'e} period, which corresponds 
in perfectly aligned graphene to Fermi levels $\sim \pm 150$ meV. 
This Fermi level scale is set by the moir\'{e} period, and can easily be reached by gating.
The substrate interaction yields particle-hole asymmetry, and in particular leads to very different densities-of-states
in conduction and valence bands.\cite{kindermann_zero-energy_2012,yankowitz_emergence_2012,wallbank_generic_2013,mucha-kruczynski_heterostructures_2013,moon_electronic_2014,dasilva_transport_2015} 

The {\em ac} conductivity is dominated by the intraband Drude peak
and interband features associated with singularities in the joint density-of-states.  
It is therefore strongly affected by the band structure modifications induced by the boron nitride substrate. 
Previous work has highlighted the feasibility of using optical absorption to determine properties of the 
substrate interaction.\cite{abergel_infrared_2013,shi_gate-dependent_2014}  
In this paper, we use a substrate interaction Hamiltonian\cite{jung_origin_2015} 
derived from {\em ab initio} electronic structure calculations 
to evaluate the {\em ac} conductivity of graphene on boron nitride.  We focus on 
the case of perfect orientational alignment. 
In contrast to other substrates for which the {\em ac} conductivity in the THz regime is dominated by a broad intraband Drude peak, 
the moir{\' e} pattern formed by graphene on boron nitride sometimes induces sharp THz peaks
due to transitions between Bloch bands formed by the moir{\' e} superlattice.  
The particle-hole asymmetry of the moir{\' e} Bloch bands 
is strongly reflected in the THz and IR conductivity which is always Drude-dominated when
the Fermi level lies above the Dirac point, but is interband-dominated 
when the Fermi energy lies in a relatively narrow interval below the Dirac point.
The qualitative change in THz and IR conductivity with Fermi energy 
suggests a potential mechanism for electrically tunable optical properties.

\noindent
{\em THz conductivity calculation:}---
We employ the moir{\'e} band Hamiltonian for graphene $\pi$-band electrons described in Ref.\onlinecite{jung_origin_2015} in 
which a local periodic substrate interaction term is added to the $\vec{k}\cdot\vec{p}$ continuum Dirac model of an isolated graphene sheet. 
The substrate interaction is extracted from {\em ab initio} electronic structure calculations\cite{jung_ab_2014,jung_origin_2015} 
and is accurate for moir{\'e} patterns with spatial periods much larger than the graphene sheet lattice constant.\cite{bistritzer_moire_2011} 
In previous work, we have used this approach to show how strains in the graphene lattice and the substrate together
with electron-electron interactions control the size of the gap at the Fermi level of neutral graphene sheets.\cite{jung_origin_2015}
We have also explained the subtle way in which substrate induced changes in carbon-site energies and hopping strengths
combine to yield surprisingly strong particle-hole asymmetry that is manifested both in the density-of-states and
in {\em dc} transport properties .\cite{dasilva_transport_2015}  

The real part of the {\em ac} conductivity can be evaluated at zero temperature and frequency $\omega$ using the 
Kubo formula expression\cite{ando_dynamical_2002,gusynin_ac_2007,falkovsky_space-time_2007,dasilva_transport_2015}
\begin{widetext}
\begin{equation}\label{eqn:sigma}
{\rm Re }\,\sigma(\omega) = \frac{\sigma(0)}{1+(\omega\tau)^{2}}+ \frac{\pi e^{2}}{\hbar}\sum_{n,\vec{k}}\frac{\Theta(\varepsilon_{F}-\varepsilon_{m,\vec{k}})-\Theta(\varepsilon_{F}-\varepsilon_{n,\vec{k}})}{\varepsilon_{n,\vec{k}}-\varepsilon_{m,\vec{k}}}\left\lvert\left\langle n,\vec{k}\left\lvert \frac{\partial H}{\partial \vec{k}}\right\rvert m,\vec{k} \right\rangle\right\rvert^{2}\delta(\hbar\omega+\varepsilon_{n,\vec{k}}-\varepsilon_{m,\vec{k}})
\end{equation}
\end{widetext}
where $\tau$ is taken to be the momentum relaxation time, $\varepsilon_{F}$ is the Fermi energy, $\varepsilon_{n,\vec{k}}$ is the energy of 
moir{\' e} band $n$ at wave vector $\vec{k}$, and $\lvert n,\vec{k}\rangle$ is the eigenstate. 
The step functions, $\Theta(x)$, ensure that transitions occur only when one state is filled and one is empty, 
while the Dirac delta function $\delta(x)$ enforces energy conservation. 
The matrix elements can be simplified by noting that the Hamiltonian depends on $\vec{k}$ only through the continuum Dirac part, $\hbar v\vec{k}\cdot\vec{\tau}$, where $\tau^{\alpha}$ are Pauli matrices ($\alpha=x,y$).

\begin{figure*}
\includegraphics[width=5in]{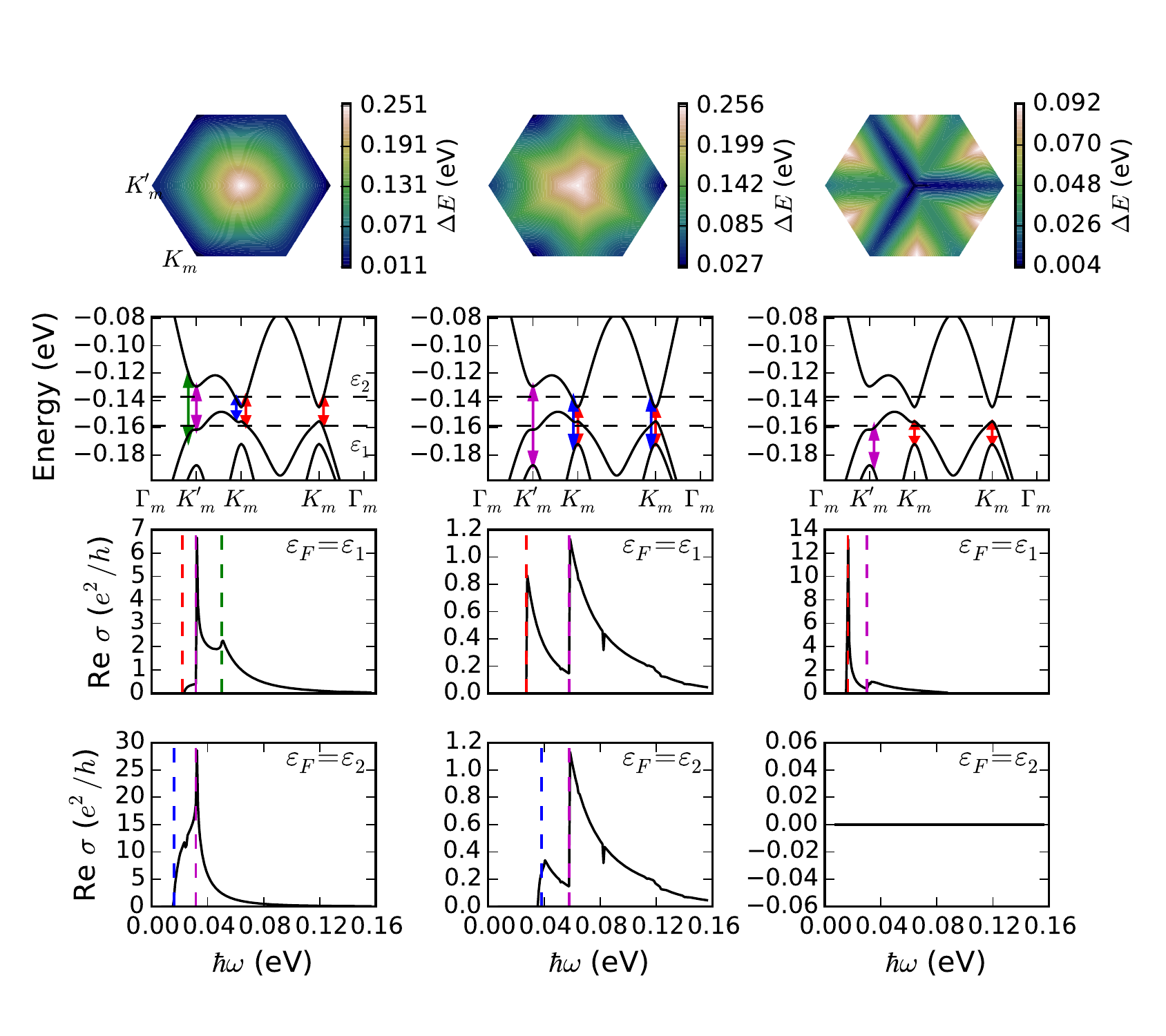}
\caption{{\bf Valence band conductivity features association with transitions at high symmetry points in the moir\'{e} Brillouin zone.} 
 The interband transition energy (first row), band structure (second row) and conductivity (third and forth rows) for interband transitions between three pairs of valence bands. The left column corresponds to transitions from the second to first valence band; the middle column corresponds to transitions from the third to first valence band; and the right column corresponds to transitions from the third to second valence band. In the band structures (second row) dominant transitions have been identified with colored arrows, and their corresponding energies are marked as vertical dashed lines on the conductivity plots (third and fourth rows). The two Fermi energies at 
which calculations were performed, marked by black horizontal dashed lines on the band structure plots, are $\varepsilon_{1}= - \hbar v G/\sqrt{3}$ and $\varepsilon_{2}= - \hbar v G/2$. The third row corresponds to Fermi energy $\varepsilon_{F}=\varepsilon_{1}$, while the fourth row corresponds to Fermi energy $\varepsilon_{F}=\varepsilon_{2}$.}\label{fig:transitions}
\end{figure*}

The results of a calculation of the {\em ac} conductivity in which Eq.~\ref{eqn:sigma} 
was evaluated numerically are illustrated in Fig.~\ref{fig:acconductivity}, which also plots the moir\'{e}
bands and summarizes the geometry of the moir\'{e} Brillouin zone.  
The plotted bands describe states near the microscopic $K$ Dirac point and given 
eigenenergy as a function of momentum measured from that point in momentum space. 
The Hamiltonian near the microscopic $K'$ Dirac point is related to the Hamiltonian at the $K$ point by time reversal symmetry; the bands and energy eigenstates are therefore degenerate in the absence of a magnetic field. 
An external magnetic field breaks the time reversal symmetry, and the {\em ac} conductivity 
contributions from the two valleys will no longer be identical. 
In this paper, we consider only the case of zero magnetic field. 

Because the moir{\' e} pattern is triangular, we choose to label symmetry points in the moir{\' e} band structure plots  
conventionally with subscript $m$, although all indicate points in momentum space close to the Brillouin-zone corner $K$ of 
the microscopic reciprocal lattice.   
The low-energy electronic structure consists of moir{\'e} bands associated with each of graphene's two valleys. 
The Kramer’s partner of a moir{\'e} band state associated with one valley lies in the opposite valley.  
For this reason the moir{\'e} band structure of a particular valley is not time reversal invariant, allowing the 
energies at moir{\'e} $K_{m}$ and $K'_{m}$ points to be different.
When the Fermi level is at the Dirac point a gap opens due to sublattice symmetry breaking. 
The conductivity is therefore zero for frequencies below the gap, peaks above the absorption 
threshold, and approaches the universal conductivity of isolated graphene
$\sigma_{0}=\pi e^{2}/2h$\cite{ando_dynamical_2002,nair_fine_2008} at high frequencies.
There is no transport gap at any Fermi level value in the conduction band,
so there is always a significant contribution to the conductivity from the Drude term peaked at $\omega=0$.
Small features superimposed on the Drude peak appear at finite $\omega$ and are associated with the 
onset of interband transition channels as the frequency increases. 
In contrast to the case in the conduction band, for clean enough samples, there is a transport gap in the valence band when the 
hole density is four per moir{\' e} peirod.   When the Fermi level lies in or just above or just below the 
moir{\' e} band edges (as shown in Figure~\ref{fig:acconductivity}) the Fermi surfaces are small, and so the 
Drude contribution remains smaller than in the conduction band case. 
Interband transitions are then dominant.  Below we associate different peaks with transitions
between moir{\' e} bands at different points in momentum space.

\noindent
{\em THz conductivity analysis:}---
In Fig.~\ref{fig:transitions} we have separated the interband conductivity into contributions from pairs of
moir{\' e} bands which cover energy intervals below the Dirac point energy. 
For each pair we have identified the dominant features, and associated them with 
transitions at or near high-symmetry points in the moir\'{e} Brillouin zone which are marked by arrows in the band structure panels
of Fig.~\ref{fig:transitions}.   First consider Fermi level $\varepsilon_{F}=\varepsilon_{1}\equiv \hbar v G/\sqrt{3}$ 
at which all three interband transitions are active.  
We order the valence bands in reverse order of energy so that the first valence band is the one closest to the Dirac point
in energy.

We see in Fig.~\ref{fig:transitions} that the second subband is flatter over a wider region of momentum 
space than either the first or the third valence band.  Accordingly third to first transitions make a smaller 
contribution to the interband conductivity than either third to second or second to first transitions.
The two strongest features are 
a third to second transition from near the $K_{m}$, $K'_{m}$ points with
energy $0.017$~eV (red in the right panel) and second to first transitions along the $K_{m}$, $M_{m}$ line with energy $0.032$~eV(purple in the left panel).  
All features can be identified with particular high symmetry points or lines 
in the moir{\' e} Brillouin zone, as shown by the matching colored arrows in the band structure plots. 
Generally speaking the strongest features are associated with interband 
transitions near the $K_{m}$ and $K'_{m}$ points in the moir\'{e} Brillouin zone.   
The top row of Figure~\ref{fig:transitions} shows differences in band energy $\Delta E$ as a function of position
in the moir\'{e} Brillouin zone as a contour plot.  
Up to a matrix element factor, the optical conductivity is proportional to the joint density of states.
For transitions from the 2nd to 1st band $\Delta E$  has a saddle point at $K_{m}$ which
leads to a divergence in the joint density of states.  In contrast, extrema in the transition energy 
yield jump discontinuities in the joint density of states\cite{daniela_dragoman_optical_2002} and weaker 
optical conductivity features.  We note that in an experiment, the presence of impurities (which we have ignored for interband transitions) will broaden the features, and lead to a finite conductivity at saddle points. 

The relatively small change in Fermi level to $\varepsilon_{F}=\varepsilon_{2}\equiv \hbar v G/2$ leads to a very different 
conductivity profile. Because both the 2nd and 3rd bands are full, transitions between these two are forbidden, leading to a zero contribution from this pair of bands. The dominant feature becomes the 2nd to 1st band transitions at the $K'_{m}$ point. This feature increases in strength as the Fermi level moves into the transport gap, and will decrease to zero as the Fermi level is moved into the first valence band. Note that this quite large change in the conductivity is seen over a relatively small change in Fermi level of $\sim 20$~meV.

For both Fermi levels, transitions from the 3rd to 1st bands are relatively weak. The reason for this is two fold. Firstly, the magnitude of the energy difference is larger. Since the conductivity is inversely proportional to this energy difference, this leads to a smaller conductivity. Secondly, as shown in the middle contour plot in Figure~\ref{fig:transitions} the transition energies at both $K_{m}$ and $K'_{m}$ for this pair of bands are minima, not saddle points, and therefore do not lead to divergences in the conductivity. The contribution to conductivity due to these transitions are jump discontinuities, and have a relatively small magnitude compared to the saddle point divergences discussed above.

\noindent
{\em Conclusions}---
We find that the conductivity of graphene on a boron nitride substrate 
is extremely sensitive to Fermi level.   The system's particle-hole asymmetry is
strongly manifested in the {\em ac} conductivity.  For Fermi levels in the conduction band, 
the Drude peak dominates at low frequencies.   For Fermi levels in the valence band, on the other hand,
interband transitions dominate when the carrier density is close to four holes per moir{\'e} unit cell.
All features in the {\em ac} conductivity are associated with transitions at high-symmetry points in the moir{\'e} Brillouin zone
which support critical points in the joint density of states.  Depending on the Fermi level, transitions between the first and second moir{\'e} valence bands, or between the second and third bands, which have transition energy saddle points and associated
divergent joint densities-of-states, dominate the total conductivity.  These large and easily tunable changes in conductivity may
be valuable for THz or IR applications.

\noindent
{\em Acknowledgements:}---
Work in Austin was supported by the Department of Energy, Office of Basic Energy Sciences under contract DE-FG02-ER45118 and by the Welch foundation under grant TBF1473. 
Work in Singapore was supported by the National Research Foundation of Singapore under its Fellowship programme (NRF-NRFF2012-01). 
We gratefully acknowledge the use of computational resources supplied by the Texas Advanced Computing Center.


%

\end{document}